\documentclass[preprint2]{aastex}
\slugcomment{Accepted for publication in APJL}
\lefthead{Dotani et al.}
\righthead{ASCA observation of SAX J1808.4--3658}
\newcommand{\sax}{SAX J1808.4--3658}

\begin{document}
\title{X-Ray Flux Decrease of the Accretion-Powered Millisecond
       Pulsar \sax\ in Quiescence detected by ASCA}

\author{Tadayasu Dotani and Kazumi Asai\altaffilmark{1}}
\affil{Institute of Space and Astronautical Science,
       3-1-1 Yoshinodai, Sagamihara, Kanagawa 229-8510, Japan}
\email{dotani@astro.isas.ac.jp}

\and
\author{Rudy Wijnands\altaffilmark{2}}
\affil{Center for Space Research, 
       Massachusetts Institute of Technology, 
       70 Vassar Street, Building 37,
       Cambridge, MA 02139-4307}

\altaffiltext{1}{Also for: Kanagawa University, faculty of engineering}
\altaffiltext{2}{Chandra Fellow}

\begin{abstract}
The accretion-powered millisecond X-ray pulsar, \sax, was observed
in quiescence with ASCA in September 1999.  We detected a dim X-ray
source in the SIS data at the position consistent to \sax.  The source
count rate was $(1.1\pm0.4)\times10^{-3}$ cts s$^{-1}$ (0.5--5 keV)
for a single SIS, which corresponds to $(3\pm1)\times10^{-14}$ ergs
s$^{-1}$ cm$^{-2}$ if a power law energy spectrum of photon index 2
with low-energy absorption corresponding to a hydrogen column density of 
$1.3\times10^{21}$ cm$^{-2}$ is assumed.
The statistical quality of the data was insufficient to constrain the energy
spectrum or to detect the 401 Hz coherent pulsation.  We compare
the data with the BeppoSAX observation also made during the quiescent
state, and find that the X-ray flux measured by ASCA is at least a
factor of 4 smaller than that measured by BeppoSAX\@.  We discuss the
possible X-ray emission mechanisms that could explain the flux change,
including the radio pulsar and the radio pulsar shock emission.

\end{abstract}


\section{Introduction}

\sax\ is a unique transient X-ray source which displays both type I
X-ray bursts and millisecond X-ray pulsations.  The source is thought
to be the formerly missing link between the neutron-star low-mass X-ray
binaries and the millisecond radio pulsars.  \sax\ was discovered in
September 1996 by the Wide Field Camera aboard BeppoSAX
\citep{zand98}.  It reached a peak intensity of 0.1 Crab and lasted
between 6 and 40 days.  Detection of two bright type I X-ray bursts
demonstrated that the compact object is a neutron star.  Recently, a
third burst was found approximately 30 days after the first two bursts
through re-analysis of a more extended set of data.  An improved
distance estimate resulted in 2.5$\pm$0.1 kpc \citep{zand00}.

The source was detected again in April 1998 with the PCA aboard RXTE
\citep{mar98}.  The observation led to the discovery of a coherent X-ray
pulsation at 401 Hz \citep{wij98}.  The pulse profile was sinusoidal
and had a low fractional rms amplitude of $\sim$4~\% (2--60 keV)\@.
Because mass accretion on to the neutron star was considered to power
the X-ray emission, a major part of the accreted mass may be channeled
to the magnetic poles without being expelled by the so-called
propeller effect.  This sets an upper limit to the surface magnetic
field of the neutron star as $(2-6)\times10^8$ G, comparable
to that of the millisecond radio pulsars \citep{wij98}.  Subsequent
analysis showed the presence of a binary period of 2.0~hr
\citep{cha98}.  This confirmed the low-mass nature of the companion.

Several transient X-ray binaries have been detected in quiescence,
and their X-ray properties were found to be quite different, depending
on the nature of the compact object in the system.  Soft X-ray
transients containing a neutron star (NS SXTs) generally have a bottom
luminosity around $10^{32}-10^{33}$ ergs s$^{-1}$ in quiescence,
whereas those containing a black hole (BH SXTs) can be as dim as or
even dimmer than $10^{31}$ ergs s$^{-1}$ \citep{asai98}.  The former
has a characteristic energy spectrum consisting of a soft component
and a hard tail.  Quiescent X-ray emission from \sax\ was discovered
with BeppoSAX on 1999 March 17--19 \citep{ste00}.  The observed MECS count
rate was very low, $\sim\!3\times10^{-3}$ cts s$^{-1}$, which may
convert to a luminosity of $\sim\!0.7-1.5\times10^{32}$ ergs s$^{-1}$
(0.5--10 keV) for an assumed distance of 2.5 kpc.  This is in the
range of quiescent luminosities in other NS SXTs.

In this letter, we report results of the 1999 September ASCA
observation of \sax\ made in the quiescent state.

\section{Observation and Results}

\sax\ was observed with ASCA from 1999 September 17, 18:10 (UT)
through September 20, 4:40, for a net exposure time of 63~ks.  ASCA
utilizes grazing-incidence, thin-foil mirrors with two kinds of focal
plane detectors: the Gas Imaging Scintillation-proportional counter
(GIS) and the Solid-state Imaging Spectrometer \citep[SIS]{tan94}.
The GIS has a wider field of view, higher time resolution, but poorer
spatial resolution and low-energy efficiency \citep{oha96}.  On the
other hand, the SIS has a higher spatial resolution, higher energy
resolution and a better low-energy efficiency.

For our quiescence observation of \sax, we had set the GIS in the
highest time resolution mode to search for the 401~Hz pulsation at the
sacrifice of the rise time information and a part of the spectral and
the spatial information.  The time resolution was 61~$\mu$s in high
telemetry bit rate and 0.5~ms in medium bit rate.  The image
resolution was reduced to $64\times64$ pixels and the total number of
spectral bins was 256.  The SIS was operated in 1-ccd faint mode
appropriate for the observations of dim sources.  The SIS data have a
time resolution of 4~s, and cover a field of view of $11'\times11'$
area.

\begin{figure*}
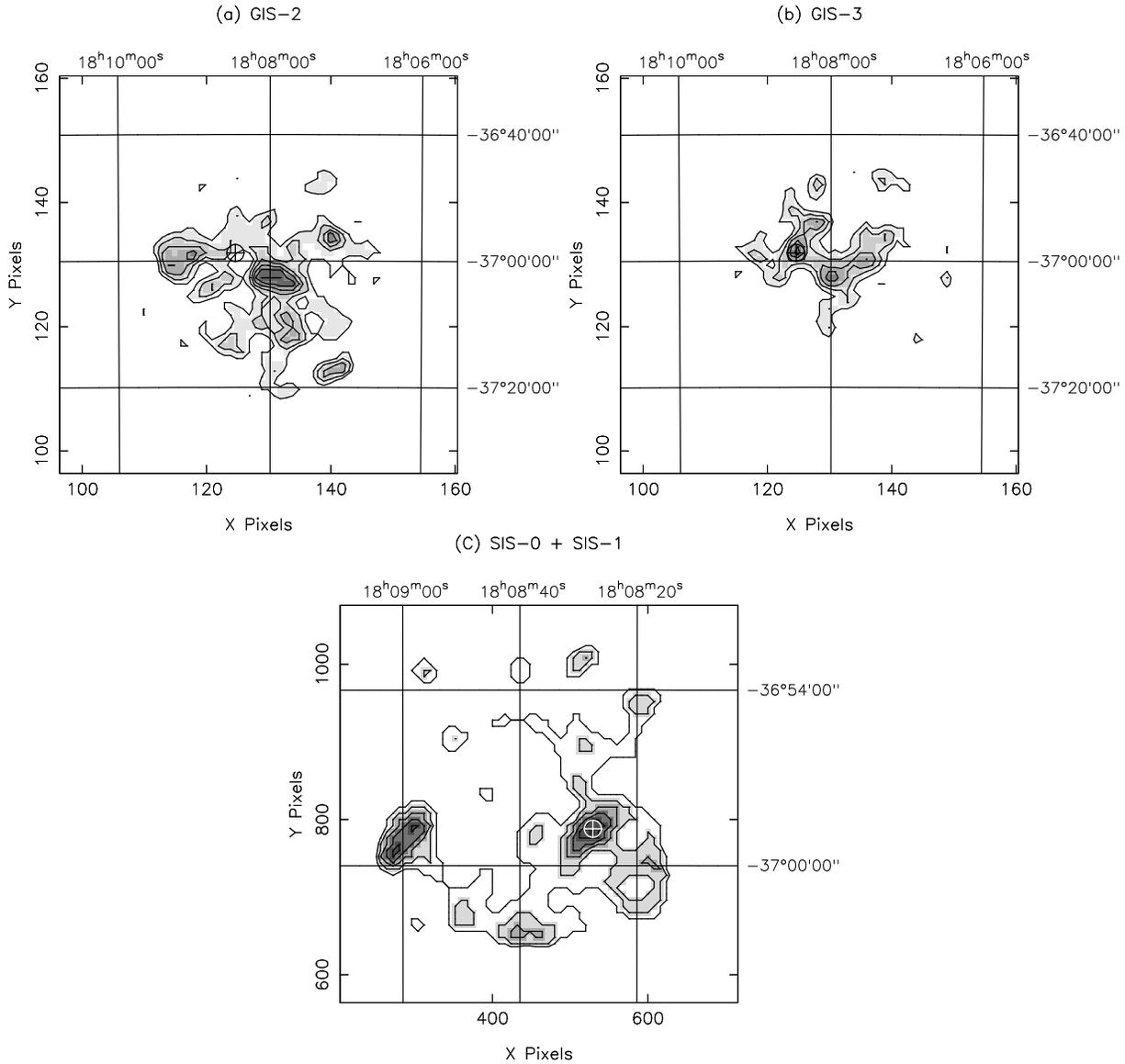

\begin{center}
\includegraphics[angle=270,width=8cm]{f1a.ps}
\includegraphics[angle=270,width=8cm]{f1b.ps}
\includegraphics[angle=270,width=8cm]{f1c.ps}
\end{center}
\figcaption{X-ray images calculated from the ASCA data; (a) GIS-2, (b)
GIS-3, and (c) sum of SIS-0 and SIS-1.  Expected position of
\protect\sax\ is indicated by the cross with a circle.  GIS images are
smoothed with a gaussian kernel of $\sigma=1'$.  Contours are plotted
in linear scale from 13.7 cts arcmin$^{-2}$ with an increment of 1.37
cts arcmin$^{-2}$.  SIS image (sum of SIS-0 and SIS-1) is smoothed
with a gaussian kernel ($\sigma=0.42'$) after rebinning by a factor of
8.  Contours are plotted in linear scale from 63.3 cts arcmin$^{-2}$
with an increment of 4.87 cts arcmin$^{-2}$.  Note that the field of
view is different between GIS and SIS.
\label{fig:x-ray_image}}
\end{figure*}

Both the SIS and the GIS data were screened with the standard data
screening criteria.  However, we could not apply the task 'gisclean',
which rejects background data using the combination of pulse height
and rise time information, because of the lack of the rise time data.
This resulted in relatively high internal background in the GIS data.
Using the screened data, we calculated an image for each sensor.  
We applied a temperature-dependent attitude correction to the SIS data,
which reduced the systematic error of the source position to 
$0.\!^{\prime}5$ (Gotthelf 1996\footnote{
http://heasarc.gsfc.nasa.gov/docs/asca/newsletters/source\_position4.html};
Ueda et al.\ 1999b).
We found that the images of SIS-0 and SIS-1 were consistent and both
showed slight excess flux at the position consistent with that of \sax.
However, the images of GIS-2 and GIS-3 look very different.  We
show in Fig.~\ref{fig:x-ray_image} the images calculated from GIS-2,
GIS-3 and summed SIS data.  The images are slightly smoothed with a
gaussian kernel appropriate for each image.  As seen in the GIS
images, the difference of diffuse emission pattern is apparent,
especially at the position of \sax; a point-like emission is seen only
in GIS-3.  Because the image difference ($\sim\!1\times10^{-3}$
c/s/cm$^2$/keV) is comparable to the local fluctuation of the internal
background of the GIS \citep{mak96}, we tried several different
screening criteria using the GIS monitor data and the satellite's
orbit related parameters.  However, the difference did not disappear.
Careful inspection of the GIS-3 image showed that the point-like
emission at the position of \sax\ actually consisted of a single
bright pixel.  Smoothing by a gaussian kernel mimicked a point-like
source.  Because the raw image does not contain such a bright pixel,
the coordinate transformation to the sky image, which includes a
position randomization process, may have produced such a bright pixel.
We confirmed that the bright pixel is not produced if we use a
difference sequence of random number.  Based on these analysis, we
consider that the coarse image binning ($1'\times1'$ bin) and high
internal background make the GIS data inadequate to search for the
faint point source. A search for the 401 Hz coherent pulsations
demonstrated also that the statistics of the GIS data were not enough
to detect the pulsations, even if the relative pulse modulation equals
100~\%.

\placefigure{fig:x-ray_image}

\begin{figure*}
\begin{center}
\includegraphics[angle=270,width=10cm]{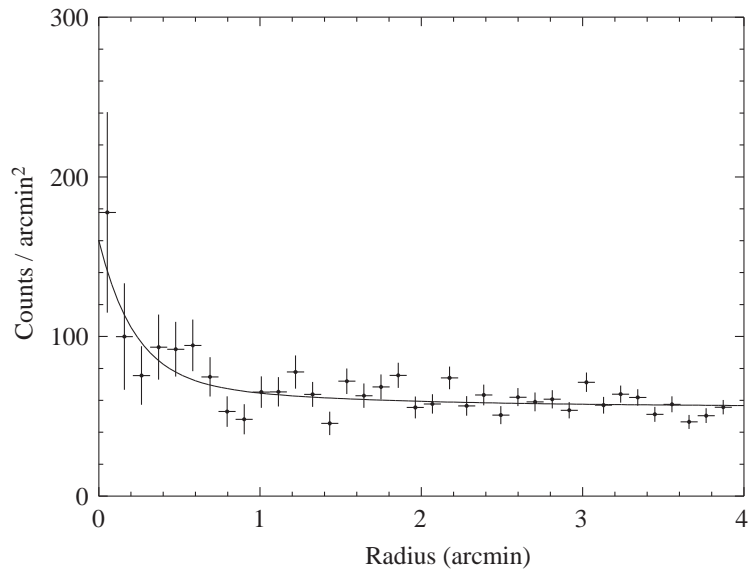}
\end{center}
\figcaption{Radial profile of the X-ray image calculated from the sum
of the SIS-0 and SIS-1 data.  The local peak of the image was taken as
the center of the radial profile.  Data points are plotted as crosses
and the best-fitting model profile for a point source (plus a constant
background) as a sold line.  No background is subtracted from the
data.\label{fig:radial_profile}}
\end{figure*}

Because the GIS data proved to be inadequate, we used the SIS data to
evaluate the X-ray flux from \sax.  We calculated a radial profile of
the source from the summed SIS image (Fig.~\ref{fig:radial_profile}).
The center of the profile was taken at the local peak of the
(smoothed) image, which coincided with the position of \sax\ within
the accuracy of position determination ($0.\!'5$ in radius).
The radial profile was
fitted with the model radial profile of a point source plus a constant
background to determine the source flux.  Precisely speaking, the
model radial profile slightly depends on the X-ray energy, although
the dependence is completely negligible compared to the statistical
errors of the present data.  The source flux was estimated to be
$(1.1\pm0.4)\times10^{-3}$ cts s$^{-1}$ (average of SIS-0 and SIS-1)
for the energy range of 0.5--5 keV\@.  Poor statistics of the data did
not allow us to extract the spectral information.  If we assume a
power law spectrum (photon index of 2) with an absorption column of
$1.3\times10^{21}$ cm$^{-2}$, which corresponds to the neutral
hydrogen column of the source direction \citep{dic90}, then the X-ray
flux would be $(3\pm1)\times10^{-14}$ ergs s$^{-1}$ cm$^{-2}$ in
0.5--5 keV\@.  This flux converts to a luminosity of $1.5-3 \times
10^{31}$ ergs s$^{-1}$ for the source distance of 2.5~kpc.

\placefigure{fig:radial_profile}

\section{Discussion}

We observed the accretion-driven millisecond X-ray pulsar \sax\ with
ASCA in quiescence.  We detected a faint X-ray source at the sky
position consistent with that of the pulsar.  However, the source was
so dim that we could not study its spectral or timing properties.  We
first elaborate on the identification of the source, and then on the
nature of the source.

Because the surface density of X-ray sources as dim as \sax\ is
relatively high, an X-ray source may happen to coincide with
\sax\ by chance.  The chance probability may be
estimated using the $\log {\rm N} - \log {\rm S}$ relation of the
X-ray sources.  ASCA survey data were used to investigate the $\log
{\rm N} - \log {\rm S}$ relation for both the cosmic X-ray background
\citep{ueda99a} and the galactic ridge emission \citep{sug00}.
According to those results, extra-galactic sources are dominated at a
flux level of $3\times10^{-14}$ erg s$^{-1}$ cm$^{-2}$.  The density of
sources brighter than $3\times10^{-14}$ erg s$^{-1}$ cm$^{-2}$ is
about $10^2$ deg$^{-2}$.  Because the ASCA position determination
error of \sax\ is about 0.5 arcmin, the probability to find a source of
$3\times10^{-14}$ erg s$^{-1}$ cm$^{-2}$ at the position consistent with
\sax\ by chance is estimated to be $\sim\!2$~\%.  
Although this probability is not very low, the source we detected is most
probably \sax.
If it were a chance coincidence of an unrelated source, 
the luminosity we calculated for \sax\ should be regarded as 
an upper limit.

The quiescent luminosity of \sax\ we obtained may be significantly
lower than that obtained by BeppoSAX \citep{ste00}.  In general, it is
necessary to assume an energy spectrum to convert the flux between
different instruments.  We assume that the spectral shape (eg.\ photon
index, absorption column) of \sax\ did not change between the ASCA and
BeppoSAX observations, and only the normalization did.  We have tried
two cases for the energy spectrum, a power law ($\Gamma = 2.0$) and a
blackbody ($kT = 0.3$ keV), which may be appropriate for NS SXTs in
quiescence.  Because the column density of \sax\ is not well
constrained, we have tried a range of the column density, from
$1.3\times10^{21}$ cm$^{-2}$ through $6\times10^{21}$ cm$^{-2}$.
Under these assumptions, the MECS count rate ($3.1\times10^{-3}$ cts
s$^{-1}$ in 1.3--10 keV) is converted to the SIS count rate using the
utility PIMMS, which is developed and maintained 
by HEASARC\footnote{Web version is available at
http://heasarc.gsfc.nasa.gov/Tools/w3pimms.html}\@.  Corresponding SIS
count rates are found to be $0.7-1.1\times10^{-2}$ cts s$^{-1}$
(0.5--5 keV) for a power law and $1.9-2.9\times10^{-2}$ cts s$^{-1}$
(0.5--5 keV) for a blackbody, respectively.  From this estimation, we
conclude that the observed SIS count rate ($1.1\pm0.4\times10^{-3}$
cts s$^{-1}$) is at least a factor of 4 smaller than the MECS count
rate, even if the statistical errors are considered.  This means that
the X-ray emission from \sax\ is time variable.

The radio pulsar shock emission, which is prefered by \citet{ste00},
may explain the time variation between the BeppoSAX and the ASCA
observations.  X-ray emission from the shocked pulsar wind has been
observed from the PSR B1259--63/SS 2883 system, which consist of a 48 ms
radio pulsar and a Be star \citep{hir99}.  ASCA observations of the PSR
B1259--63/SS 2883 system were made near periastron and apastron of the
highly elliptical orbit ($e = 0.86$) of the pulsar.  X-ray emission
was detected at both periastron and apastron, but the X-ray intensity
was different by an order of magnitude.  Because the intensity change
is well correlated to the binary phase, difference of the stellar wind
density may cause the intensity change.  If the same or a similar
mechanism works for the X-ray emission from \sax, the flux decrease
between BeppoSAX and ASCA observations may be understood as the
density decrease of the ambient matter around \sax.  However, it is
not clear what would cause the density decrease of the ambient matter.
Another emission mechanism, neutron star cooling, is also discussed by
\citet{ste00}.  Although thermal emission from the cooling neutron
star cannot explain the time variation, the mechanism may still be
valid if the X-ray emission observed by BeppoSAX was dominated by
another emission mechanism, such as residual accretion or radio
pulsar shock emission.

Although the BeppoSAX observation favors the radio pulsar shock
emission for the quiescence emission from \sax~(Stella et al.\ 2000),
another possibility can be considered for the X-ray emission detected
during the ASCA observation.  ASCA may have observed the X-ray
emission directly from the radio pulsar.  ROSAT and ASCA observations
of rotation-powered pulsars show that there is a very good correlation
between the X-ray luminosity and the spin-down energy loss rate:
$L_{\rm X} = 10^{-3} \dot{E}_{\rm spin}$ \citep{bec97}.  Here, the
X-ray luminosity, $L_{\rm X}$, is defined in 0.1--2.4~keV\@.  The
observed flux may be converted to the unabsorbed luminosity (0.1--2.4
keV) as $3.5\times10^{31} d_{2.5}^2$ ergs s$^{-1}$, where $d_{2.5}$ is
the source distance normalized by 2.5 kpc (a power law with $\Gamma =
2$ and $N_{\rm H} = 1.3\times10^{21}$ cm$^{-2}$ is assumed).  By
substituting $L_{\rm X} = 3.5\times10^{31} d_{2.5}^2$, we obtain the
spin-down energy loss rate as $\dot{E}_{\rm spin} = 3.5\times10^{34}
d_{2.5}^2$ ergs s$^{-1}$.  This leads to a surface magnetic field of
$3.8\times10^8 d_{2.5}$~G\@.  Although the distance to \sax\ is not
known accurately, this estimation of the surface magnetic field is
consistent with the limits obtained from the outburst observation
\citep{wij98,psaltis99}.  This is also consistent with the surface
magnetic field of millisecond radio pulsars.  Therefore, we consider
it possible that \sax\ might have been a radio pulsar 
during the ASCA observation.

The very low X-ray luminosity of \sax\ measured by ASCA may not fit in
the scenario that NS SXTs generally have a bottom luminosity around
$10^{32} - 10^{33}$ ergs s$^{-1}$ in quiescence, while BH SXTs do not
\citep{asai98}.  We argue that the scenario may still be valid, but
some unique feature of \sax\ makes it an exception.  Because the propeller
effect might play an important role in NS SXTs in quiescence
\citep{men99}, one may suspect that the spin period or the magnetic
field strength of the neutron star in \sax\ have very different values
than those of other neutron stars in low-mass X-ray binaries (LMXBs).
However, this is not the case.  Spin frequencies of the neutron star
may be estimated from the observations of nearly coherent oscillations
during type I X-ray bursts as observed in some neutron star X-ray
binaries or from the kHz quasi-periodic oscillations, and are mostly
found in the range 200--600 Hz \citep{kli00}.  According to the
recycle scenario \citep{bha91}, the surface magnetic field of the
neutron stars in LMXBs may be comparable to those of the millisecond
radio pulsars.  Observations of the spectral transition of
non-pulsating NS SXTs actually gave magnetic field strengths
comparable to that of the millisecond radio pulsars \citep{zha98}.
Therefore, the spin frequency and the magnetic field strength of \sax\
are not exceptional.  It is noteworthy that the outburst
luminosity of \sax\ was only $\sim\!10^{36}$ erg s$^{-1}$.
If the quiescent and
long-term average outburst luminosities are related, as is the
case if the quiescent emission is dominated by thermal emission from
the neutron star surface \citep{brown98, rutledge00}, the low
quiescent luminosity may result from the low outburst luminosity.
However, it is not clear why the outburst luminosity of \sax\ is low
(but see King 2000).   Although we do not know what makes \sax\ 
unique, the uniqueness may be responsible for both the detectable spin
modulation during the outburst and the very low luminosities in
quiescence.

\acknowledgements

RW was supported by NASA through Chandra Postdoctoral Fellowship grant
number PF9-10010 awarded by CXC, which is operated by SAO for NASA
under contract NAS8-39073.

\clearpage

\end{document}